\documentclass[12pt]{article}

\begin{document}

\centerline{\Large S.P.Novikov\footnote {Sergey P.Novikov,
 University of Maryland-College Park,
MD, 20742-2431, USA and Landau Institute of the Russian Academy of Sciences,
Moscow, 117940,
Kosygin str, 2;
 e-mail novikov@ipst.umd.edu, US tel 301-4054836(o)}}

\vspace{0.2cm}

\centerline{\bf A Note on the Real Fermionic and Bosonic quadratic
forms:} \centerline{\bf  Their Diagonalization and Topological
Interpretation\footnote{ This work is partially supported by the
NSF Grant DMS 0072700}}

\vspace{0.2cm}

{\it Abstract: We explain in this note how  real fermionic and
bosonic quadratic
forms can be
effectively diagonalized. Nothing like that exists for the general
complex hermitian forms.
Looks like this observation was missed in the Quantum Field theoretical
literature.
The present author observed  it for the case of  fermions in 1986 making
some topological work
dedicated to the problem: how to
construct Morse-type inequalities for the generic real vector fields? This
 idea also is presented
in the note.}

\vspace{0.2cm}

Let us consider a Fock space $F_n^{\pm}$  generated by the finite number
of creation operators
$a_j$ from the vacuum vector $\Phi$ such that $a_j^+\Phi=0,j=1,\ldots,n$.
 Here
we have by definition following {\bf canonical fermionic (bosonic)
 commutation
relations} between the creation and annihilation operators $a_i,a_i^+$:
$$a_i^+a_j+a_ja_i^+=\delta_{ij}$$
$$a_ia_j+a_ja_i=0,a_i^+a_j^++a_j^+a_i^+=0$$
for the space $F^-_n$ ({\bf fermions}), and
$$[a_i,a_j]=[a_i^+,a_j^+]=0$$
$$[a_i^+,a_j]=\delta_{ij}$$
for the Fock space $F^+_n$ ({\bf bosons}). Here $[a,b]=ab-ba$ in the
associative algebra
 generated by these symbols.

We introduce an euclidean  inner product $<\eta,\zeta >$ in the Fock spaces
such that $<\Phi,\Phi>=1$,
and the operators $a_i$ are adjoint to $a_i^+$:
$$<a_i\eta,\zeta>=<\eta,a_i^+\zeta>$$
According to the standard {\bf lemma} from the elementary Quantum
Field Theory,
all basic vectors in the Fock spaces $F^{\pm}_n$ have following form
$$\eta_m=a_1^{m_1}a_2^{m_2}\ldots a_n^{m_n}\Phi$$ where
$m=(m_1,m_2,\ldots,m_n)$.
We have following vectors:

$m_j=0,1$ for the case of fermions $F^-_n$

 $m_j=0,1,2,\ldots$, i.e. $m_j\in Z_+$ for bosons $F^+_n$.

 Their inner products have a form
$$<\eta_m,\eta_{m'}>=m!\delta_{m,m'}$$
where $m!=m_1!m_2!\ldots m_n!$

Therefore all these basic vectors have unit length for the case of fermions.
From the commutation relations (above) we can see that these Fock spaces are
canonically isomorphic to the spaces of the total symmetric and external powers
for the Euclidean
space $R^n$:
$$F^+_n=\sum_{k\geq 0}S^kR^n=S^*R^n$$
$$F^-_n=\sum_{k=0}^{k=n}\Lambda^kR^n=\Lambda^*R^n$$

The external power $\Lambda^*R^n$ is finite dimensional. As
it was shown by E.Witten
developing a new approach to the Morse Theory (see [1]), this presentation is
 very convenient
for the study operators acting on the spaces of differential forms in
topology.
The present author used this idea investigating the following problem:

{\bf What is a natural analog of the Morse inequalities for the singular
points of
the generic  vector fields on manifolds?}

(see Appendices to the works [2,3]; algebraic part was repeated also in the Appendix to
 the work [4]; there is a more detailed exposition of the topological part
  in [5]).
This problem led to the study real fermionic quadratic forms and their
diagonalization in [3].
It turns out that this diagonalization never appeared in the Quantum Field
theoretical
 literature before. In this note we develope  similar approach for the
  diagonalization
of the real bosonic quadratic forms. This problem also admits  a simple and
 beautiful
solution.

{\bf Definition 1}: Let us call real bosonis (fermionic) quadratic form any
self-adjoint operator $H$
acting on the Fock spaces $F^{\pm}_n$,  of the following form
$$H=U_{ij}a_i^+a_j^++V_{ij}(a_ia^+_j+a_ja^+_i)\pm U_{ij}a_ia_j+ C$$
where $U_{ij}=\pm U_{ji}$ and $V_{ij}=V_{ji}, C=const$, the signs are $(+)$
 for bosons and $(-)$
 for fermions, all coefficients are real.

This formula represents all selfadjoint operators $H$ quadratic in the
 creation-annihilation
 operators, with real coefficients.

The   diagonalization of these quadratic forms we are doing
by the following series of  steps:

{\bf Step 1.} Any real bosonic (fermionic) quadratic form can be written
in the following
{\bf Standard Real form}:
$$H=T_{ij}(a_i+a^+_i)(a_j+a^+_j))+R_{ij}(a_i-a_i^+)(a_j-a^+_j)+const$$
where $T_{ij}+R_{ij}=U_{ij},T_{ij}-R_{ij} =V_{ij},T_{ij}=T_{ji},R_{ij}=R_{ji}$
 for bosons

$$H=C_{ij}(a_i+a^+_i)(a_j-a^+_j)+const$$
where $C_{ij}=U_{ij}+V_{ij}$ for fermions.

The proof of this  is obvious.

{\bf Remark}. Nothing like that exists for the general complex hermitian
quadratic forms.

{\bf Definition 2}: By the real Bogolyubov Transformation
$a_i,a_i^+\rightarrow b_j,b_j^+$
(or canonical transformation)
we call change of basis in the linear space
$R^{2n}$ generated by the creation and annihilation operators (summation
 along the repeated indices
 is assumed here)
$$a_i=P_{ij}b_j+Q_{ij}b^+_j, a_i^+=Q_{ij}b_j+P_{ij}b_j^+$$
The new operators $b_j,b^+_j$ should satisfy to the same canonical
 bosonic (fermionic)
commutation relations; they  are adjoint to each other.

{\bf Step 2.} We prove that after the real Bogolyubov transformation
$a=Pb+Qb^+,a^+=Qb+Pb^+$
 matrix coefficients $R,T,C$
of the Standard Real forms
are changing $R\rightarrow R',T\rightarrow T',C\rightarrow C'$ according
 to the following rules:
$$T'=STS^t, R'=(S^{-1})^tR(S^{-1})$$
for bosons, where $S=P+Q$. The transformation is canonical if and only if the
following set of relations is satisfied:
$$(P+Q)(P^t-Q^t)=1$$
This transformatiom is isospectral if matrices $P\pm Q$ preserve the
orientation of the space
$R^n$. We call such Bogolyubov transformations positive.

For fermions we have
$$C'=O_+CO_-$$
where $O_{\pm}=Q\pm P$. The transformation is canonical if and only
if both matrices
$O_{\pm}$
are orthogonal. It is isospectral if Bogolyubov transformation
is positive, i.e.
$O_{\pm}\in SO_n$.

The proof of these properties can be easily obtained by the elementary
 algebraic calculation.  .

{\bf Step 3}. As a result of the step 2 we are coming to the following

{\bf Conclusion}.

I.{\it Bosons}: The symmetric matrices $T,R$ transform together as
a pair of real
quadratic forms on
the spaces $R^n$ and $R^{n*}$; beginning from now  we write $T$ as
a tensor with
 two upper indices and $R$ as
a tensor with two lower indices. We have to diagonalize them simultaneously
by the same linear transformation $S$ where $\det S>0$.
It is certainly possible if one of these forms is strictly positive
(or negative).

{\bf Theorem 1.}The diagonalization of the bosonic operator $H$ by the real
Bogolyubov transformation
is possible if and only if the matrix $(RT)_i^k=R_{ij}T^{jk}$ can be
diagonalized over the
 field $R$. It means that all eigenvalues of $RT$ should be real, and all
 Jordan cells should
be trivial. Finding the eigenbasis for the matrix $RT=(U+V)(U-V)$,
we diagonalize $H$.

Finally, we represent the operator $H$ in the form
$$H=\sum_i H_i=\sum_it_i(b_i+b^+_i)^2+r_i(b_i-b^+_i)^2, i=1,\ldots ,n$$
where $b_i,b_i^+$ are the new bosonic operators after the canonical
transformation.

In the standard model we represent $b_i,b_i^+$ by the operators
$$b=\frac{\partial + x}{\sqrt 2},
b^+=\frac{-\partial +x}{\sqrt 2}$$
 acting in the Hilbert space $L_2(R)$ of the variable $x$.
Here we have $\Phi=(const)\exp\{-x^2/2\},b^+\Phi=0$. Therefore
 the operator $H_i$
is represented by the oscillator:
$$H_i=2t_ix^2+2r_i\partial^2=(-2r_i)[-\partial^2-t_i/r_ix^2]$$
assuming that $r_i\neq 0$. We are coming to the discrete spectrum if and only if $t_i/r_i<0$
corresponding to the oscillator $-\partial^2+\omega^2x^2$ where
$\omega^2=-t_i/r_i$.
Our partial eigenvalues associated with the mode number $i$ are equal
to the numbers
$$\lambda_{i,m}=-2\frac{r_i}{|r_i|}\sqrt{-r_it_i}(m+1/2), m=0,1,2,\ldots$$
The total eigenvalues are equal to their sum
$$\lambda_{m_1,\ldots,m_n}=\sum_i \lambda_{i,m_i}$$
for all possible choices of the integers $m_i\geq 0$.

In all other cases the spectrum is continuous.

II.{\it Fermions}: The matrix $C=U+V$ transforms as
$C\rightarrow C'=O_+CO_-$ where
 $O_{\pm}$ is
a pair of orthogonal transformations. We certainly can diagonalize
 $C$ by these
 transformations. It
follows from the following  standard theorems of the linear algebra:

1.Any real matrix can be presented as a product of symmetric and orthogonal
 matrices $C=C'O_1$.

2. Any real symmetric matrix can be diagonalized by the  orthogonal
transformation $C'=O_2C'O_2^t$

3. The determinant $\det C=\lambda_1\ldots\lambda_n$ and eigenvalues
$\lambda_j^2\geq 0$ of
 the matrices $CC^t$ and $C^tC$ (the so-called ''s-numbers'')
are exactly the full set of
 invariants of the positive Bogolyubov
transformations.

{\bf Theorem 2}. The diagonalization of the fermionic operator
 $H$ by  positive
real Bogolyubov transformation
is always possible following the procedure described above
 (i.e. the diagonalization
of the matrix $C=U+V$
by the transformations $C\rightarrow O_1CO_2,O_s\in SO(n,R))$.
The eigenvalues of  fermionic quadratic form $H$ on the Fock space
$F_n^-=\Lambda^*R^n=\Lambda^{even}+\Lambda^{odd}$ have the form
$$2\sum_{l=l}^{l=k}\lambda_{i_l}-Tr C=\lambda_{i_1}+\ldots  +\lambda_{i_k}
-\lambda_{j_1}-\ldots -\lambda_{j_{n-k}}=\sum_{p=1}^{p=n}w_k
|\lambda_p|,w_k=\pm$$
where $i_1<i_2<\ldots <i_k,j_1<\ldots <j_{n-k}$ and $i_p\neq j_r$.
 The corresponding set of eigenvalues is invariant under the change of
 sign for any number of $\lambda_i$.
We permit to change sign only for even number of them, so eigenvector belongs
 to the subspace
$\Lambda^{even}$
if and only if the number of $(-)$ signs $w_k$ is even;
it belongs to $\Lambda^{odd}$ otherwise.

Let us consider now the topological problem:

{\bf What kind of Morse-type inequalities might exist for the generic real
vector fields
on the closed manifolds?}

This problem already has been discussed by the present author in 1986.
This discussion led to
the study diagonalization of the real fermionic quadratic forms.
Let us remind it here.

We consider a closed $C^{\infty}$-manifold $M_n$ with the generic
vector field $X$
(i.e. its singular points $x^*$ where $X=0$ are nondegenerate.
In the local coordinates
$y^1,\ldots,y^n$ near the point $x_j,y^k(x_j)=0$, we have
$X=(X^1(y),\ldots,X^n(y)$
and
$$X^k=C^k_iy^i+O(|y|^2),C^k_i=\partial X^k/\partial y^i (x_j)$$
In the generic case we have $\det C^k_i\neq 0$ for all singular points $x^*$.
The only local topological invariant of the vector field $X$  in the point
 $x^*$ is the sign
of this determinant $s(x^*)=\frac{\det C}{|\det C|}$. We have
$$\sum_{x^*} s(x^*)=\chi (M^n)$$
according to the Poincare'-Hopf theorem for the Euler characteristics.
Let $m_{\pm}$ are the numbers of singular points $x^*$ with the signs $\pm$
 correspondingly,
so we have $m_+-m_-=\chi (M^n)$.

{\bf Are there any separate inequalities for the numbers $m_+$ and $m_-$?}
(like the separate lower estimates).

We use for that an arbitrary Riemannian metric $g_{ij}$ on the manifold $M^n$.
It determines a 1-form $\omega$ such that $\omega_i=X^jg_{ij}$ in the local
coordinates.
We introduce a family of the operators $d_{t\omega}=d+t\omega,t\in R$,
 acting on the space of all differential
forms $\Lambda^*(M^n)$:
$$d_{t\omega}\Lambda^*=d\Lambda^*+t\omega\Lambda^*$$
Using the metric, we construct the adjoint operators $d^*$ and $\omega^*$
and the family of
second order operators
$$H_t=(d_{t\omega}+d_{t\omega}^*)^2=
-\Delta+t^2(\omega\omega^*+\omega^*\omega)+tQ$$
$$Q=\Omega +\Omega^*+d\omega^*+\omega^*d+d^*\omega+\omega^*d,
\Omega=d(\omega)$$

{\bf Lemma 1}. The $t^2$-coefficient in the operator $H_t$
 is exactly equal to the multiplication operator
by the function $<\omega,\omega>$.

The proof see below.

Therefore the zero modes $\psi_t$ of these operators $H_t\psi_t=0$ concentrate
 asymptoticaly
 $(t\rightarrow\infty)$
near the critical
points $\omega=0$ equal to the singular points $X=0$ by definition.
 The Morse-type
inequalities for
the critical numbers $m_{\pm}$ of the generic vector field $X$ are
 based on the
 fact that the
 number of
semiclassical zero modes is always not less than the number of the ''genuine''
 zero modes.

Let us calculate the number of semiclassical zero modes. Following [1,2],
we represent the space of
differential forms through the Fock spaces of fermions. Our vacuum
vector $\Phi$
corresponds to the constant function on the manifold $M^n$. We choose
orthonormal basis
in the tangent space of any given point, i.e. $g_{ij}=\delta_{ij}$ for the
 given point $x$.
Let $a^i$ be a corresponding orthonormal basis of covectors (1-forms) in the
same point.
We have $\omega=\omega_ia_i$ for every 1-form. The space of real external
forms in the point $x$
is identified with the real fermionic Fock space $F^-_n=\Lambda^*R^n$.

Proof of Lemma1: We have
$$\omega=\omega_ia_i,\omega^*=\omega_ja^+_j$$
Therefore the $t^2$-coefficient operator $\omega\omega^*+\omega^*\omega$
 has a form
$$\omega_i\omega_j(a_ia^+_j+a^+_ja_i)=
\omega_i\omega_j\delta_{ij}=<\omega,\omega>$$
Lemma 1 is proved.

For the study differential parts of these operators we choose now the
{\bf special}
local coordinates
 near the  point $x\in M^n$ such that
$$g_{ij}(x)=\delta_{ij},\frac{\partial g_{ij}}{\partial y^k}(x)=0$$

{\bf Lemma 2}. In the special coordinates associated with the point $x$
the $t$-coefficient $Q$ of the operator $H_t$
 is equal to the following expression
in the point $x$:
$$Q =U_{ij}a_ia_j+V_{ij}(a_ia^+_j+a^+_ja_i)-U_{ij}a_i^+a_j^++const=
C_{ij}(a_i+a_i^+)(a_j-a_j^+)$$
where
$$C_{ij}=\frac{\partial \omega_i}{\partial y^j}(x)$$.
 In particular, the operator $Q$ is purely
algebraic.

Proof of Lemma 2: We have $d=a_i\partial_i$ in the special coordinates
in the point $x$
where $\partial_j(a_i)=0$ in the point $x$. As a corollary we have
$Q=A+A^t+B+B^t$ where
$$A=d\omega^*+\omega^*d=
a_i\partial_i \omega_j a_j^++\omega_j a^+_j  a_i\partial_i$$
and $B$ is a multiplication operator by the 2-form $d(\omega)$:
$$B=d(\omega)=U_{ij}a_ia_j,U_{ij}=\omega_{ij}-\omega_{ji}$$
So in the point $x$ we have
$$A=\omega_{ij}(a_ia^+_j)+(a^+_ja_i+a^+_ja_i)\omega_j\partial_i=
\omega_{ij}a_ia^+_j+\omega_i\partial_i$$
 $$\omega_{ij}=\frac{\partial \omega_i}{\partial y^j}$$.

Lemma 2 follows from that immediately. Let us point out that this line
of arguments essentially
borrowed from [1], but the terms changing the number of particles did
not appeared in [1]
because
the form $\omega=df$ was exact in this work. So nontrivial Bogolyubov
transformations
did not appeared as well as in the later works of Pajitnov where the closed
 (i.e locally exact)
forms were analyzed.

As a result, we are coming to the following

{\bf Theorem 3}. For every singular point $x^*$ of the vector field $X$
the operator $H_t$
for $t\rightarrow \infty$ has exactly one semiclassical zero mode.
  It belongs to the  space $\Lambda^{even}(M^n)$
if this point is positive $s(x^*)=+$ and to the space $\Lambda^{odd}(M^n)$
if this point is negative
 $s(x^*)=-$.

{\bf Corollary}.For any closed orientable manifold $M^n$, any Riemannian
 metric $g_{ij}$
and generic vector field $X$ the numbers of the ''genuine'' zero modes for
 the operators $H_t$,
 $t$ is large enough, belonging to  the spaces $\Lambda^{even}[\Lambda^{odd}]$,
  is no more
than $m_+[m_-]$ separately.

Proof: In the quadratic approximation made in the local special coordinates
 $y^k$ near the singular
point $x^*$, we are coming to the operators
$$H_t=-\sum_i\partial_i^2+t^2C_{ij}C_{ik}y^jy^k+tC_{ij}(a_i+a_i^+)(a_j-a^+_j)
+O(|y|^3)$$
Therefore we have a potential $(C^tC)_{jk}y^jy^k$ and a fermionic quadratic
form who are
diagonalizable simultaneously by the positive Bogolyubov transformation:
 $$C^tC\rightarrow O_-^tC^tO_-=diag(\lambda_1^2,\ldots,\lambda_n^2)$$
by the rotation  of the coordinates $y\rightarrow z$,  and
$$C\rightarrow O_+CO_-=diag(\lambda_1,\ldots,\lambda_n)$$
where $\det O_{\pm}=1$.

We are coming finally to the same operators as Witten in [1]:
$$\sum_i\{-\partial_i^2+\lambda_i^2(z^i)^2+2\lambda_ia_ia^+_i\}-
\sum_i\lambda_i$$

The determinant $\det C=\lambda_1\ldots \lambda_n$ and all $\lambda_i^2$
 remain unchanged under the
 positive Bogolyubov transformation.
Therefore only even  number of the quantities $\lambda_i$ may change sign.
 The zero mode
appears here exactly like in [1] as a vector in the exterior power
corresponding to the set
of all indices where our $\lambda_i$ are negative. This number is well-defined
only modulo 2.

This argument finishes the proof.

{\bf References}.

1.E.Witten. Journ,Diff.Geometry, 1982, v 17, pp 661-692.
2.S.Novikov. Soviet Math Doklady, 1986, v 287,   pp 1321-1324
(English traslation  in vol 33(1986)). 3.S.Novikov,M.Shubin.
Soviet Mat Doklady, 1986, v 289 n 2(English translation  in v 34
(1987),n 1, pp 79-82) . 4.S.Novikov. Fields Institute
Comunications, 1999, vol 24 (dedicated to the 60th birthday of
V.Arnold), pp 397-413. 5.M.Shubin. The Gelfand Math Seminars, 1993-1995,
Birkhauser Boston, Boston, MA, 1996, pp 243-274.

\end{document}